# Tip-induced oxidation of silicene nano-ribbons

Mohamed Rachid Tchalala, [*a] Hanna Enriquez,[a] Azzedine Bendounan,[b] Andrew J. Mayne,[a] Gérald Dujardin,[a] Abdelkader Kara,[c] Mustapha Ait Ali[d] and Hamid Oughaddou [*ae]

We report on the oxidation of self-assembled silicene nanoribbons grown on the Ag (110) surface using scanning tunneling microscopy and high-resolution photoemission spectroscopy. The results show that silicene nanoribbons present a strong resistance towards oxidation using molecular oxygen. This can be overcome by increasing the electric field in the STM tunnel junction above a threshold of +2.6 V to induce oxygen dissociation and reaction. The higher reactivity of the silicene nanoribbons towards atomic oxygen is observed as expected. The HR-PES confirm these observations: even at high exposures of molecular oxygen, the Si 2p core-level peaks corresponding to pristine silicene remain dominant, reflecting a very low reactivity to molecular oxygen. Complete oxidation is obtained following exposure to high doses of atomic oxygen; the Si 2p core level peak corresponding to pristine silicene disappears.

Silicene is a new two-dimensional (2D) material with a graphene-like structure.[1] Silicene is a highly promising material because it has the advantage of being compatible with existing semiconductor technology. Substantial effort has focused on 2D silicene to discover and understand its chemical and physical properties in comparison with those already known for graphene.[2–4] Indeed, because of the similarity of the electronic properties of silicene with those of graphene, a number of theoretical studies have been made over the last few years.[1,5–7] These studies have found that the band structure of freestanding silicene exhibits a semi-metallic character,[7] and a linear dependency at the high symmetry points of the Brillouin Zone reflecting a massless Dirac fermion character.[6,7]

From an experimental point of view, a 2D silicene sheet was synthesized first on a silver surface.[8] Honeycomb structures forming either a parallel assembly of one-dimensional nano-ribbons (NRs) on Ag (100) and Ag (110)[9–13] or a highly ordered sheet of silicene on Ag (111) have been observed.[14–21] On Ag (111), at least three ordered phases were synthesized.[14–20] Recently, silicene sheets have also been grown successfully on other substrates; Ir,[22] ZrB$_2$,[23] Au,[24,25] and Pt.[26] Recent studies have also considered the formation of small Si nano-islands on the graphite surface that are quite stable.[27,28]

Even though the electronic properties of silicene are now understood reasonably well, only a few studies have reported the chemical reactivity of silicene towards hydrogenation[29] or oxidation.[30,31] Understanding the reactivity of silicene towards oxygen is of particular importance for two reasons; its stability in air and its possible integration into CMOS technology.[32] The oxidation of silicon has been studied for many decades.[33] With the advent of Scanning Probe Microscopies, it was possible to elucidate the initial reactions of oxygen at the atomic scale on many different silicon substrates.[34–36] Oxidation proceeds through dissociative adsorption via a molecular precursor.[37] In the case of silicene, only one investigation of the oxidation of isolated silicene nanoribbons (NRs) using molecular oxygen has been reported.[38] The study showed that oxidation is initiated at the ends of the Si NRs before propagating along the [−110] direction like a burning match. This indicates that the edges of the NRs are less reactive to molecular oxygen.[38] The present study is different in that here, the substrate was annealed before oxygen exposure to obtain a dense array of nanoribbons that completely cover the metal substrate surface.

In this paper, we report on the reactivity of self-assembled silicene NRs towards molecular and atomic oxygen. Insight into the electronic properties is obtained from Scanning Tunneling Microscopy (STM) and High-Resolution Photoemission Spectroscopy (HR-PES). We show that silicene NRs are less reactive to molecular oxygen compared to atomic oxygen, pointing to its relative stability toward oxygen compared to other silicon substrates. The deposition of one silicon monolayer on the Ag (110) surface induces self-assembly of silicene NRs showing a 2 × 5 LEED superstructure.[10,11] Fig. 1 presents a topographic STM image corresponding to the 2 × 5 superstructure showing self-assembled silicene nanoribbons with a very low density of defects. All silicene NRs are aligned along the Ag [−110] direction, and their length is limited only by the surface steps. The 2 × 5 rectangle indicated on Fig. 1 shows the ×5 periodicity along Ag [100] direction, with the NRs having a width of 1.6 nm. The ×2 periodicity is shown along the Ag [−110] direction. The structure of silicene nanoribbons has been extensively studied and several structures have been proposed. While studies based only on STM, and calculations[12,13,39] suggest that silicene nanoribbons have a honeycomb-like structure, others based on STM, AFM, XPD, SXRD, and DFT calculations conclude that the structure is composed of pentamers.[40–42]

[a]Université Paris-Saclay, CNRS, Institut des Sciences Moléculaires d'Orsay (ISMO), Bât. 520, 91405 Orsay, France. E-mail: Hamid.Oughaddou@u-psud.fr
[b]Synchrotron Soleil, L'Orme des Merisiers Saint-Aubin, B.P. 48, 91192 Gif-sur-Yvette Cedex, France
[c]Department of Physics, University of Central Florida, Orlando, FL 32816, USA
[d]Laboratoire de Chimie de Coordination et Catalyse, Département de Chimie, Faculté des Sciences-Semlalia, Université Cadi Ayyad, Marrakech, 40001, Morocco
[e]Département de Physique, Université de Cergy-Pontoise, 95031 Cergy-Pontoise Cedex, France

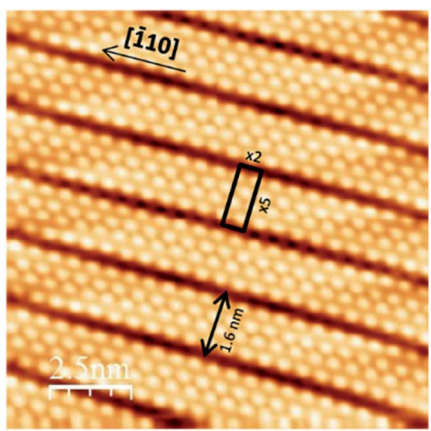

**Fig. 1** Filled-states STM image of silicene NRs grown at 200 °C on the Ag (110) surface (15 × 15 nm$^2$, $V = -0.9$ V, $I = 0.5$ nA) the (2 × 5) superstructure of the self-assembled NRs is visible inside the rectangle.

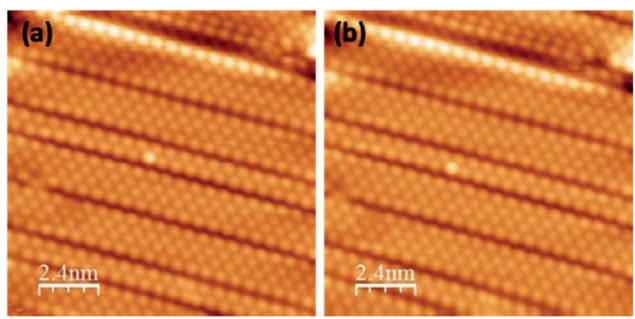

**Fig. 3** Filled states STM images of silicene NRs recorded before (a) and after (b) molecular oxygen exposure (50 L), (15 × 15 nm$^2$, −0.95 V, 0.35 nA).

Fig. 2 presents topographic STM images of the same area of silicene NRs recorded before (Fig. 2(a)) and after exposure (Fig. 2(b)) to 10 L of molecular oxygen (1 L = 1 × 10$^{-6}$ mbar per 1 s). During the exposure to oxygen, the tip was retracted about 2 microns away from the surface to avoid shadowing effects from the tip apex. After exposure to 10 L of molecular oxygen, the same small bright spots seen beforehand on the bare silicene NRs are still visible (Fig. 2(a)).

There is no apparent evidence for the adsorption of molecular oxygen with the surface, even when the dose of oxygen is increased beyond 50 L (see Fig. 3). This clearly suggests that the Si atoms do not react with molecular oxygen. No difference was observed in the STM images before and after oxygen exposure, even up to 50 L. In comparison, oxidation of Si (111)-7 × 7 occurs at very low oxygen dose (0.05 to 1 L).[34–36,43] Therefore, the silicene nanoribbons are more stable in comparison to the Si (111)-7 × 7 surface. This is also consistent with the oxidation of isolated silicene NRs, where only the ends of the silicene NRs were observed to react to molecular oxygen and not the edges.[38] Here, the dense array of nanoribbons completely covers the metal substrate surface. This reduces both the active ends of the nanoribbons and any metal activation of the oxygen reaction.

However, we were able to induce localized oxidation of the nanoribbons using the STM tip as an electron source. By positioning the STM in the middle of the image and then switching off the feedback loop, a single voltage pulse of +2.7 V, 2 nA, and 100 ms duration was applied to the surface.

The effect of a voltage pulse from the STM tip on the local structure prior to the oxidation is a necessary issue to address. Thus Fig. 4(a and b) show the impact of a voltage pulse applied to the silicene NRs in oxygen-deficient conditions. We were consistently able to find well defined and reproducible atomically resolved self-assembled NRs. Consequently, no difference is discernable between the region before and after the tip pulses and no bright protrusions appeared. We can also exclude the possibility that a direct pulse of +2.7 V on the NRs alters their uniformity or induces the formation of defects.

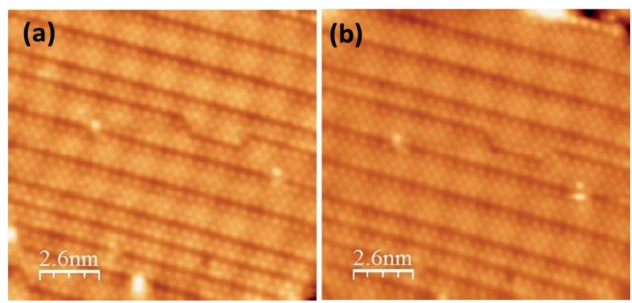

**Fig. 2** Filled states STM images of silicene NRs recorded before (a) and after (b) molecular oxygen exposure (10 L), (25 × 25 nm$^2$, −0.94 V, 0.36 nA).

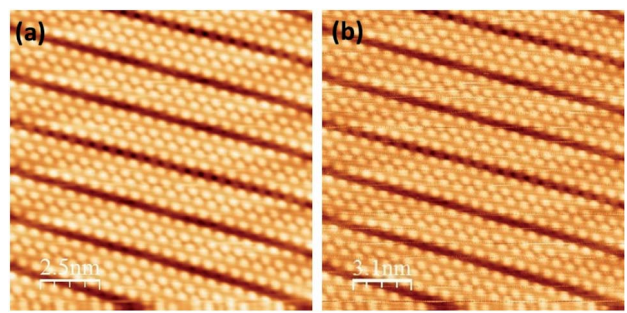

**Fig. 4** Filled states STM images of pristine silicene NRs recorded before (a) and after (b) +2.7 V pulse, (16 × 16 nm$^2$, −0.981 V, 0.482 nA).

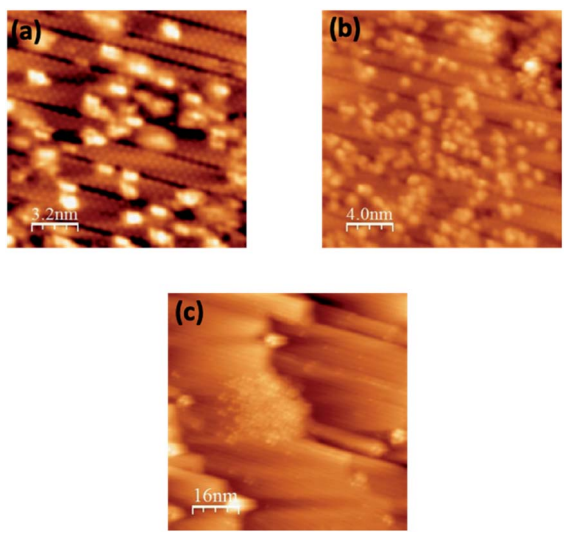

Fig. 5 Filled states STM images of silicene NRS recorded after molecular oxygen exposure followed by a tip pulse of +2.7 eV during the scan; (a) two pulses, (b) several pulses, (c) large area showing the zone where the tips pulses were applied.

Interestingly, when the same experiment was repeated after the surface had been dosed with molecular oxygen, we produced remarkably different results. In the subsequent scan, we observed the appearance of pairs of bright protrusions (Fig. 5(a)). This is a clear indication that the STM tip induces the dissociation of molecular oxygen, which then reacts with the silicene NRs. The slight variations in the angle of the bright pairs with respect to the NR indicate that there is no double tip. After the application of a second pulse, the number of bright protrusions increased. This shows that the amount of oxygen adsorbed has increased (Fig. 5(b)). Fig. 5(c) is a large STM area showing clearly the zone where pulses were performed. We applied a number of pulses using a range of voltage conditions (from +2.0 to +3.5 V). We observed a threshold for the tip-induced dissociation at around 2.6–2.7 V. Above 3 V, the dissociation process is not only very efficient but instabilities occurred regularly making a quantitative analysis difficult. When the exposure to molecular oxygen is stopped, the dissociation also stops, and no further adsorption is observed. The non-local nature of the tip-induced oxidation indicates that dissociation of the molecules is induced by the electric field between the tip and the surface. The electric field acts on the molecular dipole[44] by reducing the barrier to dissociation.[45] This has been used to create well-defined nanostructures on passivated silicon surfaces by desorption.[46,47] In contrast, an electron-attachment process would produce an atomically precise and selective dissociation[48] or desorption.[49,50] As an additional control test, pulses were applied to the bare Ag (110) surface prior to $O_2$ exposure, and no such effect on the surface was observed. This would indicate that the dissociation of

oxygen is an electronic effect induced by the electric field between the STM tip and the surface.[51]

To understand the nature of these results, we exposed the silicene NRs directly to a source of atomic oxygen. Atomic oxygen is obtained by thermal dissociation of $O_2$ with a tungsten filament held at $T = 1400$ °C. The hot filament is placed at a distance of 5 cm from the surface. Fig. 6 shows a topographic STM image recorded after exposure to 10 L of atomic oxygen on the silicene NRs. The figure clearly shows the appearance of bright protrusions similar to those observed when the voltage pulses were applied during the exposure to molecular oxygen. In addition, with this low coverage, the adsorption sites are preferentially on edge or in the middle of the ribbons. We have shown previously that molecular oxidation of isolated Si NRs propagates along the [−110] direction.[38] As mentioned before, isolated Si NRs reacted initially to molecular oxygen at their ends where there are unsaturated dangling bonds, and not the sides. Now, these 1.6 nm wide NRs are composed of two interlocked ribbons, with curved profiles.[12] Thus, reaction with atomic oxygen occurs at the outer edge atoms of the Si NRs and is favorable due to the release of strain in the NR. We also observe a tendency for the bright protrusions to line up along the [−110] direction. This can be explained by the progression of the oxidation reaction. Once the atomic oxygen has reacted with a Si atom, neighboring atoms have a modified environment with unsaturated bonds that give rise to an increase in the density of states near the Fermi level.[52] In other words, the neighboring atoms become activated, facilitating the reaction and leading to its propagation (as clusters or along the NR). This is a particularly efficient way of reducing the barrier to further oxidation[36] and has been observed during the oxidation on a number of silicon-terminated surfaces.[36,53–55] The reactivity of the Si NRs can be compared quantitatively with other silicon surfaces. The Si (111), Si (001) and SiC (001) surfaces are significantly more reactive, because of the presence of highly reactive adatoms, and Si dimers, respectively. The Si NRs are less reactive because they have electronic states further away from the Fermi level compared to the conventional silicon (100) and (111) surfaces.[56–58]

In the perspective of the reactivity of the NRs in ambient conditions, the presence of water is a crucial ingredient because it forms hydroxyl radicals (OH). These are known to play a key role in the propagation of reactions on surfaces.[58]

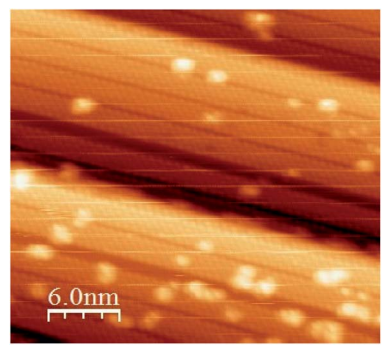

Fig. 6 Filled states STM image of silicene NRs after atomic oxygen exposure (10 L), (30 × 30 nm$^2$, −1.5 V; $I$ = 0.158 nA).

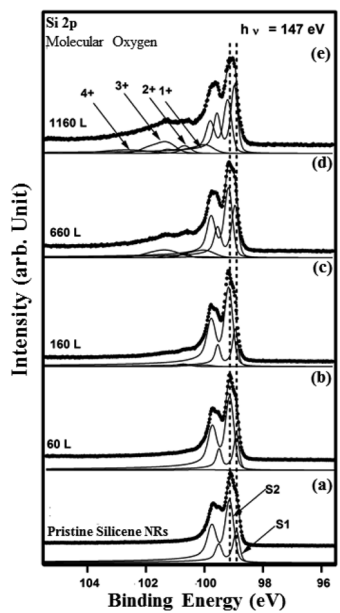

Fig. 7 Si 2p core level spectra (dots) and their deconvolutions (solid line overlapping the data points) with two asymmetric components (S1) and (S2) recorded at 147 eV for (a) the pristine (2 × 5) and ((b)–(e)) after exposure to 60 L, 160 L, 660 L, 1160 L of molecular oxygen.

To probe further the nature of the interaction between oxygen and the silicene NRs, we have performed HR-PES measurements of the Si 2p core-levels before and after exposure to molecular and atomic oxygen. The evolution of the Si 2p core level peaks with increasing oxygen exposure is displayed in Fig. 7 and 8. The spectrum corresponding to the pristine silicene nanoribbons is shown on Fig. 7(a). The Si 2p core level can be reproduced with only two spin–orbit split doublets (S1) and (S2) located at $E\ 2p_{3/2}(S1) = 98.9$ eV and $E\ 2p_{3/2}(S2) = 99.14$ eV binding energies. When we increased the amount of molecular oxygen to 60 L, no change is observed of the Si 2p (Fig. 7(b)). Additional peaks appear only when the amount of $O_2$ reaches 160 L (Fig. 7(c)).

At this exposure, only the $Si^+$ and $Si^{2+}$ oxidation states are present. At higher $O_2$ exposure (1160 L and above) additional components corresponding to $Si^{3+}$ and $Si^{4+}$ oxidation states appear at the expense of the two $Si^+$ and $Si^{2+}$ oxidation states (see Fig. 7(d) and (e)). In addition, the Si 2p corresponding to pristine silicene not only remains visible but continues to dominate the spectra. This implies that silicene NRs were not fully transformed into $SiO_2$, which is consistent with the STM results (Fig. 5(a)). The $SiO_x$ peaks are located at binding energies of +0.95 eV, +1.81 eV, +2.41 eV, and +3.65 eV corresponding respectively to: $Si^+$, $Si^{2+}$, $Si^{3+}$ and $Si^{4+}$ oxidation states of silicon. Fig. 8 shows the Si 2p core level spectra recorded on the silicene NRs before and after exposure to atomic oxygen. In contrast to the molecular oxidation results, the oxidation starts early after exposure of only 55 L of O. The components S1 and S2 corresponding to the pristine silicene decrease continuously as a function of increasing atomic oxygen dose. However, as in the case of molecular oxidation, the Si 2p corresponding to the pristine silicene is still observed even at high oxygen doses. One can notice from the PES results, a more prominent reactivity toward atomic oxygen compared to molecular oxygen. The $Si^{4+}$ component has a much higher contribution in the case of atomic oxygen after exposure to the same quantity of oxygen.

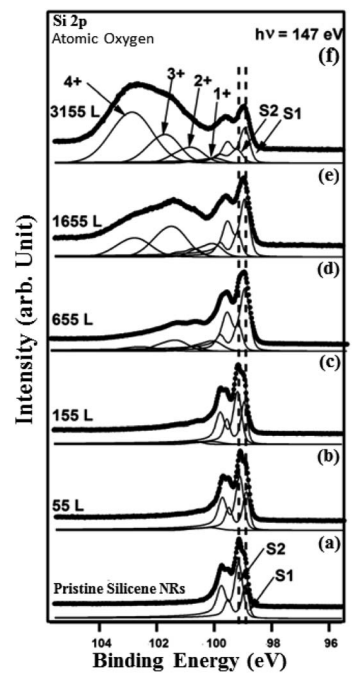

Fig. 8 Si 2p core level spectra (dots) and their deconvolutions (solid line overlapping the data points) with two asymmetric components (S1) and (S2) recorded at 147 eV for (a) the pristine (2 × 5) and ((b)–(f)) after exposure to 55 L, 155 L, 655 L, 1655 L, 3155 L of atomic oxygen.

A quantitative analysis of the XPS data confirms that silicene NRs grown on Ag (110) are much less reactive toward molecular oxygen than atomic oxygen (Fig. 9). From the plot, the percentage of $SiO_2$ increases faster during atomic oxygen exposure compared to molecular oxygen. The decrease in the percentage of clean Si remaining follows the same trend. These results agree with the conclusions from the STM images.

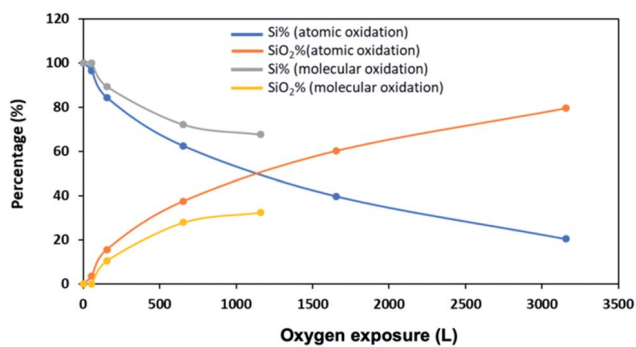

Fig. 9 Fractional intensities of silicene and $SiO_2$ versus the oxygen exposure.

## Conclusions

To summarize, scanning tunneling microscopy and photoemission spectroscopy experiments show that the functionalization of silicene with oxygen can be controlled by using either molecular or atomic oxygen. We find that silicene NRs grown on Ag (110) are much less reactive toward molecular oxygen than atomic oxygen. The silicene NRs could be oxidized in two ways. Firstly, by dissociating adsorbed molecular oxygen using STM voltage pulses, and the second, by exposing the NRs directly to an atomic oxygen flux. HR-PES data revealed the appearance of the $Si^+$ and $Si^{2+}$ components at moderately high molecular oxygen doses, while the $Si^{3+}$ and $Si^{4+}$ components only began to appear after a very large dose to molecular oxygen. The $SiO_x$ is dominated by the presence of the $Si^{4+}$ component at very high oxygen doses. Even at very high molecular oxygen doses, the Si 2p corresponding to pristine silicene NRs stays present.

## Conflicts of interest

There are no conflicts of interest to declare.

## Notes and references

1 G. G. Guzmán-Verri and L. L. Y. Voon, *Phys. Rev. B: Condens. Matter Mater. Phys.*, 2007, **76**, 075131.
2 I. Calizo, A. Balandin, W. Bao, F. Miao and C. Lau, *Nano Lett.*, 2007, **7**, 2645–2649.
3 A. C. Ferrari, J. Meyer, V. Scardaci, C. Casiraghi, M. Lazzeri, F. Mauri, S. Piscanec, D. Jiang, K. Novoselov and S. Roth, *Phys. Rev. Lett.*, 2006, **97**, 187401.
4 A. Fasolino, J. Los and M. I. Katsnelson, *Nat. Mater.*, 2007, **6**, 858.
5 A. Kara, H. Enriquez, A. P. Seitsonen, L. L. Y. Voon, S. Vizzini, B. Aufray and H. Oughaddou, *Surf. Sci. Rep.*, 2012, **67**, 1–18.
6 S. Lebegue and O. Eriksson, *Phys. Rev. B: Condens. Matter Mater. Phys.*, 2009, **79**, 115409.
7 S. Cahangirov, M. Topsakal, E. Aktürk, H. Şahin and S. Ciraci, *Phys. Rev. Lett.*, 2009, **102**, 236804.
8 B. Lalmi, H. Oughaddou, H. Enriquez, A. Kara, S. Vizzini, B. Ealet and B. Aufray, *Appl. Phys. Lett.*, 2010, **97**, 223109.
9 G. Le Lay, B. Aufray, C. Léandri, H. Oughaddou, J.-P. Biberian, P. De Padova, M. Dávila, B. Ealet and A. Kara, *Appl. Surf. Sci.*, 2009, **256**, 524–529.
10 A. Kara, C. Léandri, M. Dávila, P. De Padova, B. Ealet, H. Oughaddou, B. Aufray and G. Le Lay, *J. Supercond. Novel Magn.*, 2009, **22**, 259–263.
11 B. Aufray, A. Kara, S. Vizzini, H. Oughaddou, C. Léandri, B. Ealet and G. Le Lay, *Appl. Phys. Lett.*, 2010, **96**, 183102.
12 A. Kara, S. Vizzini, C. Leandri, B. Ealet, H. Oughaddou, B. Aufray and G. LeLay, *J. Phys.: Condens. Matter*, 2010, **22**, 045004.
13 M. R. Tchalala, H. Enriquez, A. J. Mayne, A. Kara, G. Dujardin, M. A. Ali and H. Oughaddou, *J. Phys.: Conf. Ser.*, 2014, **491**, 012002.
14 C.-L. Lin, R. Arafune, K. Kawahara, N. Tsukahara, E. Minamitani, Y. Kim, N. Takagi and M. Kawai, *Appl. Phys. Express*, 2012, **5**, 045802.
15 B. Feng, Z. Ding, S. Meng, Y. Yao, X. He, P. Cheng, L. Chen and K. Wu, *Nano Lett.*, 2012, **12**, 3507–3511.
16 C. Léandri, H. Oughaddou, B. Aufray, J. Gay, G. Le Lay, A. Ranguis and Y. Garreau, *Surf. Sci.*, 2007, **601**, 262–267.
17 Z. Majzik, M. R. Tchalala, M. Švec, P. Hapala, H. Enriquez, A. Kara, A. J. Mayne, G. Dujardin, P. Jelínek and H. Oughaddou, *J. Phys.: Condens. Matter*, 2013, **25**, 225301.
18 M. R. Tchalala, H. Enriquez, H. Yildirim, A. Kara, A. J. Mayne, G. Dujardin, M. A. Ali and H. Oughaddou, *Appl. Surf. Sci.*, 2014, **303**, 61–66.
19 H. Enriquez, A. Kara, A. J. Mayne, G. Dujardin, H. Jamgotchian, B. Aufray and H. Oughaddou, *J. Phys.: Conf. Ser.*, 2014, **491**, 012004.
20 H. Oughaddou, H. Enriquez, M. R. Tchalala, H. Yildirim, A. J. Mayne, A. Bendounan, G. Dujardin, M. A. Ali and A. Kara, *Prog. Surf. Sci.*, 2015, **90**, 46–83.
21 P. Vogt, P. De Padova, C. Quaresima, J. Avila, E. Frantzeskakis, M. C. Asensio, A. Resta, B. Ealet and G. Le Lay, *Phys. Rev. Lett.*, 2012, **108**, 155501.
22 L. Meng, Y. Wang, L. Zhang, S. Du, R. Wu, L. Li, Y. Zhang, G. Li, H. Zhou and W. A. Hofer, *Nano Lett.*, 2013, **13**, 685–690.
23 A. Fleurence, R. Friedlein, T. Ozaki, H. Kawai, Y. Wang and Y. Yamada-Takamura, *Phys. Rev. Lett.*, 2012, **108**, 245501.
24 S. Sadeddine, H. Enriquez, A. Bendounan, P. K. Das, I. Vobornik, A. Kara, A. J. Mayne, F. Sirotti, G. Dujardin and H. Oughaddou, *Sci. Rep.*, 2017, **7**, 44400.
25 M. Rachid Tchalala, H. Enriquez, A. J. Mayne, A. Kara, S. Roth, M. G. Silly, A. Bendounan, F. Sirotti, T. Greber and B. Aufray, *Appl. Phys. Lett.*, 2013, **102**, 083107.
26 M. Švec, P. Hapala, M. Ondráček, P. Merino, M. Blanco-Rey, P. Mutombo, M. Vondráček, Y. Polyak, V. Cháb and J. M. Gago, *Phys. Rev. B: Condens. Matter Mater. Phys.*, 2014, **89**, 201412.
27 M. De Crescenzi, I. Berbezier, M. Scarselli, P. Castrucci, M. Abbarchi, A. Ronda, F. Jardali, J. Park and H. Vach, *ACS Nano*, 2016, **10**, 11163–11171.
28 N. Yue, J. Myers, L. Su, W. Wang, F. Liu, R. Tsu, Y. Zhuang and Y. Zhang, *J. Semicond.*, 2019, **40**, 062001.
29 J. Qiu, H. Fu, Y. Xu, A. Oreshkin, T. Shao, H. Li, S. Meng, L. Chen and K. Wu, *Phys. Rev. Lett.*, 2015, **114**, 126101.
30 A. Molle, C. Grazianetti, D. Chiappe, E. Cinquanta, E. Cianci, G. Tallarida and M. Fanciulli, *Adv. Funct. Mater.*, 2013, **23**, 4340–4344.
31 D. Solonenko, O. Selyshchev, D. R. Zahn and P. Vogt, *Phys. Status Solidi B*, 2019, **256**, 1800432.
32 D. J. Frank, R. H. Dennard, E. Nowak, P. M. Solomon, Y. Taur and H.-S. P. Wong, *Proc. IEEE*, 2001, **89**, 259–288.
33 R. Ludeke and A. Koma, *Phys. Rev. Lett.*, 1975, **34**, 1170.
34 G. Dujardin, A. Mayne, G. Comtet, L. Hellner, M. Jamet, E. Le Goff and P. Millet, *Phys. Rev. Lett.*, 1996, **76**, 3782–3785.
35 A. J. Mayne, F. Rose, G. Comtet, L. Hellner and G. Dujardin, *Surf. Sci.*, 2003, **528**, 132–137.
36 A. Hemeryck, A. J. Mayne, N. Richard, A. Estève, Y. J. Chabal, M. Djafari Rouhani, G. Dujardin and G. Comtet, *J. Chem. Phys.*, 2007, **126**, 114707.
37 R. Martel, P. Avouris and I.-W. Lyo, *Science*, 1996, **272**, 385–388.

38 P. D. Padova, C. Leandri, S. Vizzini, C. Quaresima, P. Perfetti, B. Olivieri, H. Oughaddou, B. Aufray and G. L. Lay, *Nano Lett.*, 2008, **8**, 2299–2304.

39 P. Lagarde, M. Chorro, D. Roy and N. Trcera, *J. Phys.: Condens. Matter*, 2016, **28**, 075002.

40 G. Prévot, C. Hogan, T. Leoni, R. Bernard, E. Moyen and L. Masson, *Phys. Rev. Lett.*, 2016, **117**, 276102.

41 N. F. Kleimeier, G. Wenzel, A. J. Urban, M. R. Tchalala, H. Oughaddou, Y. Dedkov, E. Voloshina and H. Zacharias, *Phys. Chem. Chem. Phys.*, 2019, **21**, 17811–17820.

42 S. Sheng, R. Ma, J.-b. Wu, W. Li, L. Kong, X. Cong, D. Cao, W. Hu, J. Gou and J.-W. Luo, *Nano Lett.*, 2018, **18**, 2937–2942.

43 G. Comtet, G. Dujardin, L. Hellner, T. Hirayama, M. Rose, L. Philippe and M. Besnard-Ramage, *Surf. Sci.*, 1995, **331**, 370–374.

44 H. Kreuzer and L. Wang, *J. Chem. Phys.*, 1990, **93**, 6065–6069.

45 T. T. Tsong, *Phys. Rev. B: Condens. Matter Mater. Phys.*, 1991, **44**, 13703.

46 T.-C. Shen, C. Wang, G. Abeln, J. Tucker, J. W. Lyding, P. Avouris and R. Walkup, *Science*, 1995, **268**, 1590–1592.

47 L. Soukiassian, A. J. Mayne, M. Carbone and G. Dujardin, *Surf. Sci.*, 2003, **528**, 121–126.

48 B. Stipe, M. Rezaei, W. Ho, S. Gao, M. Persson and B. Lundqvist, *Phys. Rev. Lett.*, 1997, **78**, 4410.

49 P. A. Sloan and R. Palmer, *Nature*, 2005, **434**, 367.

50 L. Soukiassian, A. J. Mayne, M. Carbone and G. Dujardin, *Phys. Rev. B: Condens. Matter Mater. Phys.*, 2003, **68**, 035303.

51 J. V. Barth, T. Zambelli, J. Wintterlin and G. Ertl, *Chem. Phys. Lett.*, 1997, **270**, 152–156.

52 J. Gadzuk, *Phys. Rev. B: Condens. Matter Mater. Phys.*, 1991, **44**, 13466.

53 A. Mayne, F. Semond, G. Dujardin and P. Soukiassian, *Phys. Rev. B: Condens. Matter Mater. Phys.*, 1998, **57**, 15108.

54 G. Dujardin, A. J. Mayne and F. Rose, *Phys. Rev. Lett.*, 1999, **82**, 3448–3451.

55 F. Amy, H. Enriquez, P. Soukiassian, P.-F. Storino, Y. Chabal, A. Mayne, G. Dujardin, Y. Hwu and C. Brylinski, *Phys. Rev. Lett.*, 2001, **86**, 4342–4345.

56 J. Wintterlin, R. Schuster and G. Ertl, *Phys. Rev. Lett.*, 1996, **77**, 123–126.

57 B. Stipe, M. Rezaei and W. Ho, *Science*, 1998, **279**, 1907–1909.

58 C. Sachs, M. Hildebrand, S. Völkening, J. Wintterlin and G. Ertl, *J. Chem. Phys.*, 2002, **116**, 5759–5773.